\documentclass[10pt]{article}
\usepackage{graphics,epsfig}
\usepackage{harvard}
\usepackage{amsmath,amssymb,amsthm}
\usepackage[tocindex]{apacite}
\usepackage{url}

\newcommand{\bibnodot}[1]{}

\begin{document}

\title{Quantum Theory and Conceptuality: Matter, Stories, \\ Sematics and Space-Time}
\author{Diederik Aerts\\
        \normalsize\itshape
        Center Leo Apostel for Interdisciplinary Studies \\
        \normalsize\itshape
        Vrije Universiteit Brussel, 1160 Brussels, 
       Belgium \\
        \normalsize
        E-Mails: \textsf{diraerts@vub.ac.be}}
\date{}
\maketitle

\begin{abstract}
\noindent We elaborate the new interpretation of quantum theory that we recently proposed, according to which quantum particles are considered conceptual entities mediating between pieces of ordinary matter which are considered to act as memory structures for them. Our aim is to identify what is the equivalent for the human cognitive realm of what physical space-time is for the realm of quantum particles and ordinary matter. For this purpose, we identify the notion of `story' as the equivalent within the human cognitive realm of what ordinary matter is in the physical quantum realm, and analyze the role played by the logical connectives of disjunction and conjunction with respect to the notion of locality. Similarly to what we have done in earlier investigations on this new quantum interpretation, we use the specific cognitive environment of the World-Wide Web to elucidate the comparisons we make between the human cognitive realm and the physical quantum realm.
\end{abstract} 

\begin{quotation}
\noindent
\end{quotation}

\section{Introduction} 
The general aim of this article is to continue the elaboration of the new interpretation of quantum mechanics that we presented in Aerts (2009b, 2010a,b). Our focus in this article is to acquire a deeper insight into the similarities and differences between the human cognitive realm and the realm of quantum particles as conceptual entities with respect to the notions of matter and space-time. In previous articles we expressed the view that the human cognitive realm is still much less organized as a conceptual structure than the quantum cognitive realm (Aerts 2009b, section 4). Our reflections about the notions of matter and space-time attempt to make this difference more concrete and also to identify in more depth still the fundamental similarities.

In our new interpretation of quantum theory (Aerts 2009b, 2010a,b), quantum entities are mediating as conceptual entities between pieces of ordinary matter that function as a memory structure for these quantum entities. By ordinary matter we mean substance made of elementary fermions, i.e. quarks, electrons and neutrinos, hence including all nuclei, atoms, molecules, macroscopic material objects and also measuring apparatuses. Human concepts and combinations of them, i.e. sentences, pieces of text, etc \ldots, are mediating between human minds or artificial memories. Most plausibly due to the billions of years of evolutionary fine-tuning, by means of selection and variation, the quantum mediating cognitive conceptual process has acquired a very deep structural symmetry. This is why it can mathematically be adequately modeled by means of the quantum formalism, as it exists now, which, although quantum theory as a physical theory is very complex, is in essence a relatively simple mathematical structure. In this quantum formalism, states of quantum entities are represented by unit vectors of a complex infinite dimensional Hilbert space, and observables linked to measurement processes are described by self-adjoint operators on this complex Hilbert space. The evolution dynamics is described by Schr\"odinger equation, or more generally by a unitary transformation of the Hilbert space. The measurement dynamics is described by orthogonal projection operators of the spectral decomposition of the self-adjoint operator representing the observable to be measured. As a consequence of a measurement, the state is projected and normalized, which in the quantum jargon is called collapse.

The human mediating cognitive process is only thousands of years old, and hence still in a very primitive stage of development as compared to the quantum mediating process. This means that any mathematical theory for the human cognitive realm should be expected to be much more complex than the quantum formalism in a non-trivial way. However, because of the deep ontologic correspondence -- both are fundamental cognitive processes between memory structures -- the quantum formalism can be applied to describe and model quite a number of the effects appearing in human cognition, as has been shown by the numerous results obtained in the newly emerging domain called `quantum cognition' (Aerts 2009a, Aerts and Aerts 1995, Aerts, Aerts and Gabora 2009, Aerts and Czachor 2004, Aerts and D'Hooghe 2009, Aerts and Gabora 2005a,b, Aerts, Gabora, Sozzo and Veloz 2011, Bruza and Cole 2005, Bruza, Kitto, McEvoy and McEvoy 2008, Bruza, Kitto, Nelson and McEvoy 2009, Busemeyer, Wang and Townsend 2006, Busemeyer, Pothos, Franco and Trueblood 2011, Gabora and Aerts 2002, Khrennikov and Haven 2009, Pothos and Busemeyer 2009, Van Rijsbergen 2004, Widdows 2003, Widdows and Peters 2003). This means that the quantum formalism can certainly serve as a basis for the development of a powerful mathematical formalism for human cognition. 

In reflecting on the structure of the human cognitive process that we put forward in this article, we will pay special attention to similarities with the quantum formalism. This involves the risk that, like the quantum structure itself, we may be aiming at a structure that is too simple and concrete already to capture all of the human cognitive process. More in-depth research will therefore need to be conducted in the future than what we are able to present here and have presented in previous articles (Aerts 2009b, 2010a,b). Such research could be inspired by the State Concept Property (SCOP) formalism (Aerts 2002, Aerts and Gabora 2005a,b, Gabora and Aerts 2002) that we developed in earlier studies of the use of the quantum formalism to model the combination of human concepts.
 
\section{Abstract, Concrete, Concepts and Objects}
There are two parts of analysis regarding the new quantum interpretation that we have performed on previous occasions and that in this article will guide us in identifying the structure that plays the role with respect to human cognition that matter and space-time play with respect to the quantum mechanical realm.

The first part of analysis is widely manifest in earlier publications on the new interpretation (Aerts 2009b, Aerts 2010a,b). It is presented in great detail in the analysis of how the Heisenberg uncertainty relation is encountered in the human cognition realm in section 4.1 of Aerts (2009b). The more abstract a human concept is, the less concrete it is, and vice versa, and this is the expression of Heisenberg's uncertainty for the case of human concepts. For example, the concept {\it Cat}, without any specification, is a rather abstract concept, whereas if we consider {\it This Cat Felix}, and we mean `this particular and unique cat named Felix, the one I can touch and caress with my hand', then this is a most concrete form of the concept Cat. On several occasions, we introduced the notion of `state of a concept'. According to this notion, the most abstract version {\it Cat} and {\it This Cat Felix} each represent a state of {\it Cat}. Hence, for each concept there are states corresponding to more abstract forms of the concept and states corresponding to more concrete forms. Heisenberg's uncertainty principle for human concepts expresses that a concept cannot be at once in a very concrete and in a very abstract state. This is an expression about the ontological nature of what concepts as mediating entities can be. In our new interpretation of quantum theory, this is also the way Heisenberg's uncertainty principle is explained for quantum entities. A quantum entity cannot be in a very concrete state -- a state close to being a localized state -- and in a very abstract state -- a state close to being a state of definite momentum -- at once. On several occasions, we have also introduced the conceptual environment of the World-Wide Web to provide examples and explanations and we will do this again in the present article. 

Suppose that we google for the word `cat'. On September 5, 2011, this returned 2,330,000,000 hits, which means that, on that day, there were 2,330,000,000 webpages listing the word `cat' at least once. In the conceptual environment of the World-Wide Web, the totality of combinations of concepts contained in each of these webpages constitutes also a state of the concept {\it Cat}, where all other concepts in this combination are conceptual contexts that change the most abstract state of {\it Cat} to the most concrete state for this specific conceptual environment. Indeed, the conceptual content of webpages containing the word {\it Cat} are the most concrete states of {\it Cat} if we consider the World-Wide Web as our specific conceptual environment. Of course, each one of these most concrete states of {\it Cat} is also a most concrete state of many other concepts, namely the concepts appearing in the text contained in the relevant webpage. It is in this sense that if we focus on the conceptual environment which is the World-Wide Web, we may consider the collection of all webpages, more specifically their conceptual content, as the analogue for the case of human concepts of what the content of space is for the case of quantum particles. More concretely, if one of the webpages is chosen, opened on a computer screen, and looked at by a person, this is the analogue for the case of human concepts of what a snapshot of space and its content, hence localized states of different quantum entities looked at by an observer, is for the case of quantum particles. The current level of order and structure of the collection of webpages of the World-Wide Web is far from that of the collection of quantum particles structured in entities of ordinary matter or in fields of bosonic nature, available to appear as a snapshot of localized states in space. On a fundamental level, however, the similarity can be identified.

In quantum theory, a localized state of a quantum particle is complementary to a momentum state, i.e. a state where the momentum of the particle is localized in momentum space, and the Heisenberg uncertainty principle stands for the incompatibility of both types of state, i.e. for a quantum particle there are no states that are strongly localized in position space and strongly localized in momentum space. The more abstract the form of a concept, the more it is incompatible with a very concrete form of the same concept. Hence the collection of all abstract forms of human concepts, for example the collection of words in a dictionary, correspond with the snapshots of momentum space and its content. These abstract forms of
concepts are the analogue of quantum particles with well-determined momentum, but almost completely non-localized in position space. Let us take a concrete situation to make this clear. This time we consider the conceptual environment of human memories. The most concrete state of a concept then is the state it has in a specific human memory, where the context is defined by all aspects of this human memory. If two persons communicate with each other by means of the spoken word, then strings of abstract forms of concepts are sent from one human memory to another human memory, triggering these concepts stored in memory, changing their states, or exciting them. The resulting dynamics is what we refer to as communication between two human minds. When quantum particles emitted by a radiating piece of ordinary matter hit another piece of ordinary matter, atoms or molecules in this piece of matter get excited and, when de-exciting, will send out again quantum particles that can eventually be captured by the original piece of ordinary matter. This is a typical situation of matter interacting with quantum particles, and hence also matter interacting through quantum particles with other matter, or matter communicating with matter.

The second part of analysis is linked to the fundamental difference between a concept and an object. We have reflected about this in several sections of our previous articles, particularly in section 5 of Aerts (2010b). For a concept $A$ and a concept $B$, we have that `$A$ or $B$' is again a concept. However, if $A$ and $B$ are objects, then `$A$ or $B$' is not an object. A `chair or table' is not an object but a concept. With respect to the logical connective `and', we do not encounter this fundamental difference between a concept and an object. Indeed, if $A$ and $B$ are concepts, then `$A$ and $B$' is a concept, but also if $A$ and $B$ are objects, then `$A$ and $B$' is an object. Remark that if $A$ and $B$ are physical objects, hence objects that both occupy a part of physical space, then the object `$A$ and $B$' will occupy both parts of physical space, hence it occupies a part of physical space which is the set theoretic union of the parts of physical space occupied by $A$ and $B$ if we consider physical space as a set of points. The foregoing observation contains an important hint with respect to the identification of the human cognition equivalent of what physical space is in the case of quantum particles.

If we consider classical logic and the conceptual combinations which are called propositions, then with respect to such propositions there is a complete mathematical symmetry of the logical connectives `and' and `or'. This symmetry is reflected in the set theoretic model of classical logic, where `and' corresponds to the operation of `intersection' or `meet' and `or' to the operation of `union' or `join'. But also in the formation process of human concepts in itself, the connectives `and' and `or' play a very symmetric role. In Aerts (2009a) section 4.1, we considered this formation process, and we will briefly return to the insight presented there. Through the process of `concept formation', the two connectives, disjunction and conjunction, play an equally important role. Consider for example the concept {\it Animal}. {\it Animal} can be {\it Dog} or {\it Cat} or {\it Horse} or {\it Rabbit} or \ldots followed by a long list of all the usually known animals. Hence {\it Animal} is a typical example of a concept where disjunction has played a fundamental role in its formation. Conjunction can play an equally fundamental role. Consider as an example the concept {\it Dog}. Then the conceptual combinations {\it Has Four Legs} and {\it Likes to Bark} and {\it Has Fur} and {\it Likes to Swim} and \ldots followed by a long list of characteristics of a {\it Dog}, play an essential role in the formation of the concept {\it Dog}. In the realm where `objects' are considered, the connective `or' drops, and we remain with `and' alone. Also, the connective `and' has acquired an intense relation with the notion of space as theatre where `objects can take place'.

Both parts of analysis -- the first, connecting abstraction and concretization with Heisenberg uncertainty and considering the World-Wide Web as an example of a cognitive environment where the most concrete states of concepts are the conceptual contents of webpages where these concepts appear, and the second, analyzing the fundamental difference between concepts and objects, and how the connectives `or' and `and' behave in this respect -- are the guidance of the hypothesis that we want to put forward in the next section and that will put us on the trail of identifying the equivalence for human cognition of what matter and physical space-time are for quantum particles.

\section{Human Cognitive Proto-Matter}
To identify the equivalent for the human cognitive realm of what physical space-time is for physical reality, we need to investigate first what the equivalent is for the human cognitive realm of what ordinary matter is for physical reality. In previous writings on our new interpretation of quantum theory, we have indicated `human memory' or `an artificial memory system capable of interaction with human semantics' as the equivalent for the human cognitive realm of what ordinary matter is for physical reality. In the present article we will elaborate this in more detail.

Following the dual process theory of cognition, two types of human thought are distinguished (Barrett, Tugade and Engle 2004, Bruner 1990, Freud 1899, James 1910, Kahneman 2003, Paivio 2007, Sloman 1996, Sun 2002). Type 1 thought is largely unconscious, automatic, contextual, emotional and speedy. Type 2 thought is deliberate, explicit, effortful and intentional. It turns out that most of human behavior is shaped by the inarticulate type 1 thoughts. 

In earlier work we also identified two modes of thought (Aerts and D'Hooghe 2009) inspired by the mathematical structure of the quantum modeling scheme we developed for human concepts (Aerts 2009a), and called them `quantum conceptual thought' and `classical logical thought'. Without doubt there is a correspondence between the two types of thought from dual process theory and the two modes of thought we introduced in Aerts and D'Hooghe (2009). The correspondence, however, is not necessarily a morphism, also because while dual process theory relies on experimental evidence and on theoretical hypothesis related to different aspects of human cognition, our classification finds its origin in the mathematical structure of the quantum modeling scheme. Roughly, what we have called quantum conceptual thought in a comparison would correspond to type 1 thought and what we have called classical logical thought would correspond to type 2 thought. We have planned to investigate in depth the nature of this correspondence in future research, but here we will merely use a specific aspect of type 1 thought, which is the following. When a human subject is confronted with a stimulus, it is commonly so that type 1 thought quite spontaneously gives rise to a story or at least a fragment of a story, such that different elements of this stimulus `fit the story'. It is this `story fitting' aspects of human type 1 thought that interest us particularly in what follows in this section.

Let us give an example of what we mean. Suppose we consider an experiment where the stimulus consists of words on a screen shown to the participants in the experiments. Consider more specifically the situation where the stimulus is the word `bank'. Experiments show indeed that even a stimulus consisting of one word, such as `bank', is enough to give rise to a story for a participant in the experiment due to type 1 thought. In the case of the word `bank', the story might be about money. Or it can be even more concrete, containing an image of the building of a known bank and a visit to this bank. Or the story can be about a conversation with a staff member of this bank, etc \ldots. We deliberately choose as an example the word `bank', because it is a word with several meanings. Suppose that one of the individuals participating in the experiment is a fervent fisher, then the word `bank' may well evoke `the bank he or she sits on while fishing', giving rise to quite a different type of story. For example, `what happened the last time he or she went fishing', etc \ldots. If the stimulus consists of the two words `bank' and `money', it is most likely that only a story containing the first meaning of `bank' is produced, whereas a stimulus consisting of the two words `bank' and `fishing' will most probably trigger only a story containing the second meaning of `bank'. Hence, on presentation of a stimulus consisting of only the word `bank', there may be a very brief instant in which the participant's mind vacilates between the different meanings and their very different associations. Experiments suggest that such a state of ambiguity rapidly resolves towards one of the two stories under type 1 thought. If we consider this event within our classification of quantum conceptual thought, we would say that the state of the concept {\it Bank} collapses to being part of one of the two stories.

Before we give a more specific description of the equivalent for the human cognitive realm of what matter is for physical reality, we want to make a specific observation on ambiguity and stories. We mentioned already that within the process of type 1 thought in case there is ambiguity about which of the stories will best fit the stimulus data, type 1 thought functions in such a way that the ambiguity is resolved rather quickly, and one of the stories is elected. This means that the connective `or', if appearing to express ambiguity between two or more stories pertaining to different meanings of the stimulus, is removed by type 1 thought. Of course, we have not explicitly made clear what is meant by `different meanings of the stimulus'. However, what we want to show in the following is that the choice within type 1 thought to optimize the removal of ambiguity, has a deep influence on the nature of the stories that we allow as entities. Let us make this more concrete.

If we consider one story $A$ and a second story $B$, then the story `$A$ and $B$' is again a story. In an extreme case, when there is no meaningful connection between $A$ and $B$, the new story `$A$ and $B$' is nothing more than two separated stories $A$ and $B$, but this we still consider as a story, albeit of an extreme type. In most cases, however, there will spontaneously emerge meaningful connections between $A$ and $B$, such that `$A$ and $B$' is a new story which is more than the two stories $A$ and $B$ separately. Indeed, it would be rare for two stories $A$ and $B$ not to contain meaningful connections in any individual's life such that they merge spontaneously to a third story `$A$ and $B$'. For two stories $A$ and $B$, the cognitive construction `$A$ or $B$' is usually not considered to be a story. Looked at from a purely conceptual point of view -- i.e. if we consider a story just as a combination of concepts -- then `$A$ or $B$' is again a combination of concepts, and hence again a story. But if we put the connective `or' between two stories, although in theory this gives rise to a story, it will usually not be considered as a story, because the ambiguity is introduced in an artificial way, so that its reduction is not optimized, taking into account the global meaning landscape of the human mind involved. Let us go back to the example of the stimulus `bank'. It is very well possible that this stimulus gives rise to ambiguity in the mind of a participant in the experiment, invoking a story consisting of `$A$ or $B$' -- where in story $A$, the word `bank' is associated with `money', and in story $B$, with a place where you can `sit'. But if such ambiguity appears in the `$A$ or $B$' story, it will be considered `an ambiguity to be removed', which is why `$A$ or $B$' will usually not be considered a story.

On many occasions, we have taken the World-Wide Web as a possible cognitive environment, primarily because it allowed us to collect experimental data by making use of search engines (Aerts 2009b, Aerts 2010a,b, Aerts 2011, Aerts, Czachor, D'Hooghe and Sozzo 2010). We will take the same approach in this article, and this time the webpages will play the role of what we have called stories. Using the Yahoo search engine, let us show how indeed the connectives `and' and `or' play different roles in webpages of the World-Wide Web. On September 15, 2011, we found the word `and' to return to 1,610,000,000 Yahoo hits, and the word `or' to return 5,400,000,000 Yahoo hits. This means that `or' appears more often than `and' on the World-Wide Web, although both frequencies of appearance are of the same order of magnitude, and their proportion is 1,610,000,000/5,400,000,000=0.3.

We then elected two words that had no obvious connection, viz. the words `car' and `building'. The number of Yahoo hits for `car and building' was 8,450 and the number of Yahoo hits for `car or building' was 7,810 -- we carried out searches for the appearance of the expressions `car and building' and `car or building' in their entirety, hence by entering double quotation marks on both sides of the expression in a Yahoo search engine. To compare these frequencies of appearance systematically, let us introduce
\begin{equation}
C({\rm car \dots building})={N({\rm car\ and\ building}) \over N({\rm car\ or\ building})}
\end{equation}
where $N({\rm car\ and\ building})$ is the number of webpages containing the part of sentence `car and building' and $N({\rm car\ or\ building})$ the number of webpages containing the part of sentence `car or building'. Hence, we have
\begin{equation}
C({\rm car \dots building})={8,450 \over 7,810}=1.08
\end{equation}
If we use longer combinations that carry more meaning, such as `the car and the building', we get 2,950 Yahoo hits, while `the car or the building' returns 33 hits, which means that the proportion has increased to 89, indeed we have $C$(the car \ldots the building)=2,950/33=89. 

We consider a second example using the two words `flute' and `bass'. We have `flute and bass' giving rise to 11,900 Yahoo hits, while `flute or bass' to 162, hence a proportion of 73.4. If we look at a longer part of sentence including the two words `flute' and `bass', we find for `the flute and the bass' 68 Yahoo hits, and `the flute or the bass' 1 Yahoo hit, hence a proportion of 68. For the next example, we consider the two words `horse' and `house'. For the part of sentence `horse and house' we find 12,500 Yahoo hits, and for the part of sentence `horse or house' we find 4,690 Yahoo hits, hence a proportion of 2.6. The longer part of sentence `the horse and the house' gives rise to 73 Yahoo hits, while `the horse or the house' gives rise to 5 Yahoo hits, hence a proportion of 14.6. Table \ref{orandandcomparison} presents the different examples and their respective proportions, and we will analyze the results in the following.
\begin{table}[h]
\caption{A systematic comparison between the frequency of appearance of the connectives `and' and `or'}
\begin{center}
\begin{tabular}{|cccc|}
\hline
expression & `and' hits & `or' hits & proportion \\
\hline
& 1,610,000,000 & 5,400,000,000 & \underline{0.3} \\
car \ldots building & 8,450 & 7,810 & 1.1 \\ 
the car \ldots the building & 2,950 & 33 & 89.4 \\
flute \ldots bass & 11,900 & 162 & 73.4 \\
the flute \ldots the bass & 68 & 1 & 68 \\
horse \ldots house & 12,500 & 4,690 & 2.6 \\
the horse \ldots the house & 73 & 5 & 14.6 \\
table \ldots sun & 8,900 & 123 & 72.4 \\
the table \ldots the sun & 83 & 3 & 27.7 \\
\hline
window \ldots door & 4,090,000 & 937,00 & 4.3 \\
the window \ldots the door & 9,000 & 21,900 & \underline{0.4} \\
the window \ldots door & 61,900 & 22,800 & 2.7 \\
laugh \ldots cry & 297,000 & 779,000 & \underline{0.4} \\
to laugh \ldots to cry & 11,100 & 11,400 & \underline{1} \\
to laugh \ldots cry & 31,400 & 311,000 & \underline{0.1} \\
dead \ldots alive & 149,000 & 13,100,000 & \underline{0.01} \\
being dead \ldots alive & 3,270 & 9,010 & \underline{0.3} \\
wanted dead \ldots alive & 47,100 & 2,240,000 & \underline{0.02} \\
coffee \ldots tea & 2,860,000 & 3,690,000 & \underline{0.7} \\
drinking coffee \ldots tea & 8,580 & 26,800 & \underline{0.3} \\
wants coffee \ldots tea & 2 & 92 & \underline{0.02} \\
want coffee \ldots tea & 51 & 8,230 & \underline{0.006} \\
milk \ldots sugar & 1,510,000 & 24,600 & 61.3 \\
wants milk \ldots sugar & 10 & 4 & 2.5 \\
want milk \ldots sugar & 141 & 179 & \underline{0.8} \\
\hline 
\end{tabular}
\end{center}
\label{orandandcomparison}
\end{table}

It should be noted that the World-Wide Web is still far too small to provide significant statistics for longer parts of sentence than the ones we have considered. Indeed, a part of sentence such as `the red car and the high building' already returns zero hits, as does the part of sentence `the red car or the high building'. However, we predict that once the World-Wide Web has grown to the extent that searches for long sentences, and eventually even paragraphs return substantial numbers of pages containing these longer sentences or paragraphs, the proportion between the connective `and' and the connective `or' will increase for long parts of combinations of concepts, where the combinations are made with concepts chosen without obvious connection. However, if the occurrence of `and' as a connective in sentences is more frequent than that of `or', why are there three to four times more single `or' connectives than `and' connectives, the numbers of Yahoo hits being 1,610,000,000 for the connective `and' and 5,400,000,000 for the connective `or'. Could it be that there is a mistake in how Yahoo counts these pages? There is not, and the following examples explain why. Indeed, the state of affairs that we are bound to detect and that explains why there is no mistake, will also lead us to the identification of the proto structure of matter within the human conceptual realm. 

So, for the next example, we will consider the two words `window' and `door'. For `window and door' and `window or door', we find 4,090,000 hits and 937,000 hits, respectively, hence a proportion of 4.3, still an increased frequency of the connective `and' as compared to the connective `or'. Next we consider the part of sentence `the window and the door', which gives 9,000 hits, while the part of sentence `the window or the door' gives 21,900 hits. This suddenly inverses the proportion, i.e. for this part of sentence, the frequency of the connective `or' is higher than that of the connective `and'. The proportion of `and' to `or' is 0.4. Let us try to understand this phenomenon by looking at some specific webpages that appear in the Yahoo search. For example, when searching for the part of sentence `the window or the door', we found webpages where it appeared in the phrase `Do you prefer your bed facing the window or the door to your room?', and in `Easily mounted by adhesive tape to the window or the door', and again `But Holmes credits himself for quickly adapting and revising his theory once he was personally convinced that no danger could enter the room from the window or the door'. When we searched for the part of sentence, `the window and the door', we found that the first webpages all contained the sentence `Hidden behind the window and the door', followed by several webpages containing the sentence `Close the window and the door'. The inversion of the proportion means that the part of sentence `the window or the door' is more frequent in the meaning structure of human cognition than the piece of sentence `the window and the door'. There is another aspect we need to point out. Let us consider the part of text `the window and door', for which Yahoo gives 61,900 hits, against 22,800 hits for `the window or door'. This means that for these very similar parts of text the proportion between `and' and `or' is normalized again, namely 2.7. In short, it is for the parts of text `the window and the door' and `the window or the door' that the inversion takes place.

Let us examine another example to understand better this phenomenon. We considered the two words `laugh' and `cry'. For the parts of text `laugh and cry' and `laugh or cry', we found 297,000 and 779,000 Yahoo hits, respectively, which means again an inversion of the same order of magnitude as the one we identified for `window' and `door', i.e. 0.4. We then considered the parts of text `to laugh and to cry', which yielded 11,100 hits, and `to laugh or to cry', giving 11,400 hits, i.e. a proportion equal to 1. This means that the inversion disappeared again. Lastly, we considered the parts of text `to laugh and cry', with 31,400 Yahoo hits, and `to laugh or cry', with 311,000 Yahoo hits, giving a proportion of 0.1, which indicates a very strong supremacy of the connective `or' over the connective `and' for this part of sentence. 

Hence, in the case of `window' and `door', it is the specific part of sentence `the window or the door' which introduces a strong weight with respect to the appearance of the ambiguity connected to the connective `or', while in the case of `laugh' and `cry', it is the specific part of sentence `laugh or cry' which introduces a strong weight with respect to this ambiguity introduced by the connective `or'.

Let us consider a third example, namely the combinations of the two substantives `dead' and `alive' using the connectives `and' and `or' . The combinations are `dead and alive', with 149,000 hits, and `dead or alive', with 13,100,000 hits, hence a proportion of 0.01, which is one tenth of what we found earlier. Here, the inversion is enormous. For `being dead and alive' we found 3,270 hits, and for `being dead or alive', we found 9,010 hits. This means that the effect of inversion almost disappeared when the combination of concepts `dead or alive' was entered in the part of text including the concept `being' in front, the proportion being 0.3. But if we consider `wanted dead and alive', with 47,100 hits, and `wanted dead or alive', with 2,240,000, the proportion is 0.02, which is again of the order of magnitude of the expression itself.

The next example concerns the words `coffee' and `tea'. For `coffee and tea', we found 2,860,000 hits, and for `coffee or tea', we found 3,690,000 hits, i.e. a proportion of 0.7. When we put the word `drinking' in front, however, the change was substantial. We found that `drinking coffee and tea' returned 8,580 hits, against 26,800 hits for `drinking coffee or tea', hence a proportion of 0.3. We then tried several more combinations. We entered `wants coffee and tea', with 2 hits, and `wants coffee or tea' with 92 hits, a proportion of 0.02, which is the order of magnitude we found for `dead' and `alive'. We also entered `want coffee and tea', with 51 hits, and `want coffee or tea', with 8,230 hits, a proportion of 0.006, the smallest we had found so far.

Our final example considers the words `milk' and `sugar'. For `milk and sugar', we found 1,510,000 hits, and for `milk or sugar', we found 24,600 hits, i.e. a proportion of 61.3, the order of magnitude of the biggest results we found so far. Furthermore, we searched for `wants milk and sugar', with 10 hits, and `wants milk or sugar', with 4 hits, hence a proportion of 2.5. For `want milk and sugar', the number of hits was 141, and for `want milk or sugar', the number of hits was 179, hence a proportion of 0.8.
 
When we combine two concepts that we have chosen more or less at random, such as {\it Car} and {\it Building}, {\it Flute} and {\it Bass}, {\it Horse} and {\it House}, and {\it Table} and {\it Sun}, Yahoo searches of the corresponding words `car' and `building', `flute' and `bass', `horse' and `house', and `table' and `sun' on the World-Wide Web indicate that combinations with the connective {\it And} in between these concepts are more common than combinations with the connective {\it Or}. The connective {\it Or} introduces an abstraction, and, taking into account our identification of the Heisenberg uncertainty as related to abstraction and concretization (Aerts 2009b, section 4.1), this means that where the connective {\it Or} is substituted in between two concepts (Aerts 2009b), a superposition state is formed, which is less localized than the two component states. On the contrary, in general, the connective {\it And} introduces a concretization. This means that where the connective {\it And} is substituted in between two concepts, a pure state which is more localized is formed. Since our experimentation with the World-Wide Web shows that, for randomly chosen concepts, the longer the combination, the more common the {\it And} connective becomes as compared to the {\it Or} connective, this indicates the general tendency towards localization of texts to be found on webpages of the World-Wide Web. This process towards localization stops per definition at the cognitive end-products, which are the concrete webpages contained in the World-Wide Web. If we take the World-Wide-Web as an example of a cognitive environment, it is these concrete webpages that are the equivalents for human cognition of what ordinary matter is for physical reality. Within a classical vision on physical reality, it is believed that matter fills up space-time by giving rise to objects. In previous articles, e.g. Aerts (2009b) section 4.3, we already analyzed why this classical vision is the limit of a process towards objectivation, where, however, the status of object as such is never reached. The notion of object is therefore only an idealized notion playing a valuable role in the idealized theory which classical physics is. This is borne out by physical ordinary matter, which is never really localized, because it contains atoms and molecules, and inside of these substructures particles are in superposition states which are not local. Hence, we encounter a similar situation explicitly in the realm of human cognition. The {\it Or} connective, giving rise to non-localized states, consistently appears in large numbers in the form of small `molecules of meaning' in the webpages of the World-Wide Web. The examples we identified are {\it The Window Or The Door}, {\it Laugh Or Cry}, {\it Dead Or Alive} and {\it Coffee Or Tea}. These are the equivalents for the human cognitive realm of what the molecules and atoms of ordinary matter are for physical reality. 

There is another thing we wish to point out. At first sight, it might seem that the molecules of meaning of the human cognitive realm are immobile, as if nothing moved inside them, in apparent contrast with the highly dynamic nature of atoms and molecules of ordinary matter, with electrons moving around nuclei made up of protons and neutrons. First of all, the idea of `electrons moving around a nucleus', something like a miniature solar system, is an image that we know to be very wrong. It is an image that, again, is forced upon us because quantum particles are presented as tiny ping pong balls bumping and bouncing around (Aerts 2010b). Most text books on quantum physics state rather explicitly that the image of the tiny solar system is wrong, because the electrons `move in a cloud around the nucleus'. Although this is supposed to rectify the prevailing erroneous idea of quantum particles, the resulting image is very wrong yet again. There is nothing that really moves within a molecule or an atom. One step towards a better definition would be to say that `the electrons are in a cloud around the nucleus', and that `this cloud changes as time elapses'. Of course, the expression `as time elapses' should also be specified with care. It actually means `as time elapses when measured in a laboratory where experiments are performed with molecules and atoms'. With respect to this time, the cloud of presence of the electrons changes. And even this is not correct. It is not `a cloud of presence', but `a cloud of potential presence'. And, if we add the word potential, the word cloud is in fact no longer correct. In short, the following statement would be much closer to being correct, `Electrons are in states of potential presence, and this presence -- which is `not' actualized in general -- is situated around the nucleus. And it is the potential which changes as time in the laboratory elapses. Let us consider the molecule of human cognition, `coffee or tea'. The typical situation that we can imagine with respect to this molecule of human cognition is the following. At a reception for a specific event, coffee and tea are served. One of the visitors of the event is presented a tray with cups of hot coffee and cups of hot tea, with the person holding the tray uttering the words `coffee or tea'. Let us now zoom in on the mind of the visitor, who likes both of the drinks offered. Before making a choice, the visitor is most likely to see the different alternatives pass before his or her eyes. Literally, this means that the potentialities with respect to the coffee versus tea choice are changing as the time within visitor's mind elapses. This `change of potentialities of the coffee or tea alternative' is the equivalent for the human cognitive realm of the change taking place in a molecule in physical reality. It is one of the research aims of our Brussels research group to work out a concrete model for this, but, most of all due to so many other fascinating research aims we are working on at the moment, we have not yet had the opportunity to do so in very explicit terms. Over a decade ago, however, we succeeded in elaborating a model of change within the human cognitive realm, more specifically for the situation of the liar paradox (Aerts, Broekaert and Smets 1999a,b), and the approach and method used for this dynamical model of the liar paradox can readily be used for a description of the dynamics of the `coffee or tea' cognitive molecule. Something of this nature was done, for example, for the situation of the Prisoner's Dilemma situation by Jerome Busemeyer and collaborators (Busemeyer, Wang and Townsend 2006, Busemeyer, Pothos, Franco and Trueblood 2011, Pothos and Busemeyer 2009). 

\section{Human Cognitive Reality and Physical Space-Time}
Physical space-time is the theatre of ordinary matter. More specifically, it is the imagined place and time where snapshots filled with ordinary matter interacting with quantum particles can be situated. Hence, the equivalent of this physical space-time for the human cognitive realm is the theatre of stories. It is, by the way, interesting to remark that the word `story' is derived from the Latin `historia' and the Greek `$\iota\sigma\tau o \rho\iota\alpha$', which in turn is derived from the Proto-Indo-European root `weid-'. This root has given rise to the following derivations in different languages: English `ywis', English `iwis', English `wise', English `wisdom', English `witan', English `wite', French `guise', Greek `eidos', Greek `Haides', Greek `histor', Irish `find', Latin `videre', Provencal `guidar', Sanskrit `vedah'. 

How and where can stories be situated? Again, the example of the World-Wide Web can help us to gain a better insight into the structure that reveals itself by identifying stories as the equivalent for human cognition of ordinary matter for physical reality. If we consider each webpage, or interconnected website, as a story, the collection of all stories then becomes the collection of all webpages, which is the World-Wide Web. Can we identify a space-time like structure connected to the World-Wide Web? It is quite obvious that no space-time like structure very similar to physical space-time can be identified connected to the World-Wide Web. But then, there is no need for that, because, like we mentioned already, we expect the structures connected to the human cognitive realm to be much more complex than the equivalent structures connected to the quantum realm. And there is a whole body of scientific research that is of value with respect to this question, even from before the World-Wide Web existed. Indeed, scientists have extensively studied the semantic structure of large bodies of texts, and also proposed mathematical models for it, called `semantic spaces'. The core of most of these semantic analysis approaches is the so called `document-term matrix', which contains as entries the number of times that a specific term appears in a specific document. Suppose that we label the rows of the matrix by the documents and the columns by the terms, then each row of the matrix can be seen as a vector representing the corresponding document, and each column as a vector representing the corresponding term. If vectors are normalized, the scalar product amongst such normalized vectors is a measure of the similarity of the corresponding documents and terms, and it is also used as such in theories of information retrieval and semantic analysis. In the vector space of vectors representing terms, the documents are represented by the canonical base vectors of this vector space. This means that also the similarity between terms and documents can be calculated by means of the scalar product of the corresponding vectors, and in this way documents can be compared with search terms, and the most relevant documents can be taken to be the most similar ones. This is more or less how today's search engines on the World-Wide Web work, although in practice there are many variations on this basic approach. Vector space models for semantic analysis and information retrieval were first introduced by Salton, Wong and Yang (1975). Recent examples of such approaches are Latent Semantic Analysis (LSA) (Deerwester et al. 1990), Hyperspace Analogue to Language (HAL) (Lund and Burgess 1996), Probabilistic Latent Semantic Analysis (pLSA) (Hofmann 1999), Latent Dirichlet Allocation (Blei, Ng and Jordan 2003), or Word Association Space (WAS) (Griths and Steyvers 2002). Connections with quantum structures have been investigated from different perspectives within the previously mentioned emergent domain of research called `Quantum Cognition' (Aerts and Czachor 2004, Arafat and van Rijsbergen 2007, Van Rijsbergen 2004, Widdows 2003, 2006, 2008, 2009, Widdows and Peters 2003).

Let us have a closer look at LSA (Deerwester et al. 1990), for which we analyzed correspondences with quantum physics in Aerts and Czachor (2004). LSA explicitly introduces rank lowering of the document-term matrix by considering the singular value decomposition of this matrix and substituting some of the lower singular values by zero. One reason for introducing this rank lowering technique is to render the sparse matrix of very high rank into a less sparse matrix of less high rank, which makes it easier to manipulate from a mathematical point of view. There is also an effect of de-noisification, since the original document-term matrix is noisy due to the presence of anecdotal instances of terms. However, there are two more subtle aspects that are of specific interest to our analysis. If some of the lower singular values are substituted by zero, and the approximated document-term matrix is calculated, it can be shown that the places where the original document-term matrix had zeros, because the terms did not appear in the document, will now contain numbers different from zero. This means that the new document-term matrix reveals `latent' connections between documents and terms. Even if a term does not appear in a specific document, but does appear in many documents similar to this document, the matrix will contain a number different from zero for this term and this document, expressing that, although the term does not appear in the document, it is relevant for the document. Another aspect is that the terms of the square matrix appearing after the singular value decomposition can be interpreted as `conceptual dimensions'. These terms indeed correspond in some way to `directions of strong relationships between the terms and documents', and if we express these directions conceptually, they can be interpreted as `conceptual dimensions'. Even analyses of small samples using the LSA technique may produce high numbers of these dimensions. This is an expression of what we mentioned already, namely that the human cognitive realm is still much less organized than physical reality, where quantum particles interact with ordinary matter. For this realm of physical reality, three space dimensions have shown to be able to grasp all of the structure, at least on the macroscopic level.

To date, LSA has proved one of the most powerful semantic analysis formalisms. The procedures are fully automatic and allow to have texts analyzed by computers without any involvement of human understanding. LSA produced particularly impressive results in experiments with simulation of human performance. LSA-programmed machines were able to pass multiple-choice exams such as a Test of English as a Foreign Language (TOEFL) (after training on general English) (Landauer and Dumais 1997) or, after learning from an introductory psychology textbook, a final exam for psychology students (Landauer, Foltz and Laham 1998). LSA certainly owes much of its potential to its ability to calculate the similarity between a term and a document without the need for the term to appear in the document. The mathematical technique penetrates the meaning structure which is at the origin of the texts to be found in the documents, which are only snapshots of this meaning structure. Hence, by introducing a non-operational mathematical ingredient, the lowering of dimension by means of singular value decomposition and dropping of lower singular values, the LSA approach manages to introduce a mathematical description that is a better model of the underlying meaning structure.

Since the World-Wide Web is a large collection of texts, the semantic space approaches can also be applied directly to it, which is what search engines do. If words are typed into a search engine, the pages of the World-Wide Web which are `closest' to these words are gathered and presented to the individual that is doing the search. How the notion of `closest' is calculated depends on the type of semantic space taken as a foundation of the web search engine, and possibly on other aspects of relevance. Anyhow, `closest', and hence also `close', `less close', `further away', `far away' and `farthest' are estimations that can be calculated numerically within such a semantic space model of the World-Wide Web, and they are always linked to `meaning'. It is possible to define a `meaning bond' directly on the World-Wide Web (Aerts 2011), and identify aspects of concept combinations such as the guppy effect by using this meaning bond (Aerts, Czachor, D'Hooghe and Sozzo 2010).

Even less so than is the case for the World-Wide Web or other large bodies of text, the collection of human stories as the proto-matter of human cognition will have an easily identifiable semantic structure, although some of the problems encountered for the World-Wide Web or other large bodies of text are not, or at least less, present for the collection of human stories. For example, unlike human minds, search engines need to work with `words' and cannot directly work with `concepts'. Equally so, a story is different from a collection of words in that it is also a conceptual entity.
Hence, to develop the mathematical structure of human cognition, it is possible to focus on `concepts' rather than on `words', and on `the conceptual entities that stories are' rather than on `the bag of words that a webpage is'.

Although it is a very important and intriguing problem to find out what is the most adequate topological and/or metric structure of meaning within the realm of human cognition, hence how concepts and stories can be mathematically represented such that their intrinsic connections are modeled, in the next part of this section we want to focus on the global insights into physical reality we can infer from our identification of the equivalent for human cognition of what ordinary matter and physical space is for human cognition. Indeed, independently of the topological and/or metric structures, structural elements can be identified on a more profound level.

Let us again consider the World-Wide Web as our working example. We will also make use of the operational analysis we have elaborated for space-time and relativity in earlier work (Aerts 1996a,b, 1999). This analysis carefully distinguishes between the different elements that are underlying the reality of space-time, taking into account the insights gained through operational quantum theory within the Geneva-Brussels approach (Piron 1976, 1990, Aerts 1982, 1983), and hence introducing explicitly a role for `the effect of measurement' and `the construction aspect of elements of reality' also in relativity theory. The main elements of this operational analysis are the following.

We consider the following situation. An observer $O_1$ has a specific experience $e_1$ which is his or her `present experience' at a particular moment of time, which we call $t_1$, measured by his or her watch. This `present experience' contains only a tiny part of the reality that exists at this moment $t_1$ for this observer. How can we know in an operational way what is the rest of the existing reality at moment $t_1$ for this observer $O_1$? We propose the following operational procedure, borrowed from quantum theory. At some moment in the observer's past, he or she could have made a decision such that his or her present experience, hence the experience at time $t_1$, is different from $e_1$, for example $e_1'$. Also, the part of reality contained in experience $e_1'$ exists for the observer at time $t_1$. At another moment in the observer's past, yet another decision could have been made leading up to a third experience $e_1''$ at time $t_1$. Also the reality contained in this third potential experience $e_1''$ exists at time $t_1$ for the observer. Hence, if we call ${\cal E}$ the set of all potential experiences that the observer could have lived at time $t_1$ if he or she would have made decisions in the past leading to one of these experiences, then the reality contained in each one of these experiences exists at time $t_1$ for the considered observer $O_1$. In Aerts (1996a,b), we showed that if relativity theory is interpreted geometrically, namely that the length contraction and time dilation effects calculated in relativity theory are real space-time shifts and not physical effects on rods and clocks, future events in some reference frame are contained in the present reality at time $t_1$ of the observer $O_1$ that we consider. The reason is that the considered observer could have decided in the past to go and travel close to the speed of light so that, on his or her return, time on earth would have elapsed much more than the time indicated on the observer's watch. This means that at time $t_1$ future events in earth time are real for the observer. Hence, reality is four-dimensional, containing, in addition to the three dimensions of space, also a dimension that reaches out into the future in this specific way. We analyzed this situation in detail in Aerts (1996a,b), and refer to these articles for the subtleties involved. Although the analysis presented in these earlier papers proves that there is no paradox involved, it still is a situation that is difficult to grasp, of course. We will now show that our new interpretation of quantum mechanics sheds new light on this situation too.

Let us once more consider the World-Wide Web as a cognitive environment of the human cognitive realm, and analyze the reality-time structure that emerges. Observing now consists in electing a webpage of the World-Wide Web and reading this webpage. Let us consider an observer $O_1$. We start by considering experience $e_1$ taking place at time $t_1$ on the observer's watch, while the experience consists of electing website $w_1$ and reading it. The reality contained in this present experience is webpage $w_1$, more specifically the meaning content of webpage $w_1$. But in the observer's past, he or she could have made another decision, such that at time $t_1$ another webpage $w_1'$ would have been elected, and the experience $e_1'$ would take place, consisting in electing this webpage $w_1'$ and reading it. This means that also the meaning content of webpage $w_1'$ is part of the reality of observer $O_1$ at time $t_1$. The same line of reasoning can be followed for all webpages that can be elected and consulted by the observer. Let us make the hypothesis that all existing webpages are available to be elected by an observer. Hence, as a consequence the semantic content of the collection of all existing webpages is the reality at time $t_1$ for this observer $O_1$. This conclusion still fairly well corresponds to what we would intuitively think to be the `semantic reality' at time $t_1$ of the observer $O_1$, since he or she can, if he or she wants to, indeed elect and consult any of the existing webpages. Let us consider a second observer $O_2$. For the first observer $O_1$, the semantic content of all webpages is `real' at time $t_1$, `because' he or she could elect any of this content and experience it at time $t_1$. For a second observer $O_2$, we can follow an equivalent line of reasoning, and hence as a consequence the semantic content of all of the webpages is real at time $t_2$ for this observer, where $t_2$ is a time measured on the watch of the second observer $O_2$. This means that for both observers $O_1$ and $O_2$, the whole semantic content of the World-Wide Web is their reality at any moment of time on their respective watches. Whenever they act by making a part of this global semantic content of the whole World-Wide Web into their present experience, they elect a parcel of place-time reality and the stories contained in it in the human cognitive realm. By `place-time' we mean `semantic place' and time, where the semantic place is defined by a semantic theory, for example one of the semantic spaces we mentioned in the foregoing section.

Let us apply the insight gained in the above to physical reality and physical space-time. Hence we start from the basic hypothesis of our new quantum interpretation, namely that, fundamentally, in the realm of the physical reality of ordinary matter interacting with quantum particles a similar state of affairs exists as in a process which is a conceptual communication process. If this hypothesis is true, the role that the observer played in our foregoing analysis within the human cognitive realm is now played by pieces of ordinary matter within the realm of physical reality. These pieces of matter communicate with each other by means of quantum particles. In the course of these communication processes, place-times are elected where these processes `take place (and time)'. Of course, since our human body is itself such a piece of matter, it participates in these processes whenever we as humans and as physical entities are confronted with these processes of communication between pieces of matter through quantum particles. To avoid any confusion, it should be noted that here we `do not' participate with our human minds in the human cognitive realm. Or again, `we do not speak with these material entities'. Although we see them, i.e. participate in these processes by means of photons, we do not speak with them. This is why, to our human mind, which is an entity interaction in the human cognitive realm, this communication happening in the physical realm is interpreted as `the experiencing of snapshots of space-time filled with objects made of matter'. This is a wrong interpretation. We imaginarily paste together all these snapshots of space-time to a space-time continuum and picture for ourselves the situation as if pieces of matter were moving around in this space-time continuum as material objects. This erroneous interpretation originated classical mechanics and it was not until the advent of quantum theory that its faulty nature could be pointed out. Let us remark that relativistic effects such as time dilation can be naturally explained by the analysis presented in this article. Indeed, there is no space-time filled with objects consisting of matter. On the contrary, it is matter interacting with other matter through quantum particles that time and again locally gives rise to a place-time parcel, i.e. a space-time snapshot. Exactly like -- to return to the realm of human cognition -- the webpage elected at a certain time gives rise to the place-time parcel, this time within semantic space, where the semantic interaction can be localized. The reason why the locally brought about space-time snapshots hang together to form a relatively smooth global space-time continuum for the global reality is because all these local snapshots are indeed grounded in one reality, which, however, is not inside a space-time. Again, we can understand this phenomenon by comparing it with how it happens in the human cognitive realm. All the locally elected webpages hang together such that they can be looked at as forming a relatively smooth global body of text, because all of them are grounded in one reality, namely the reality of global human knowledge.

\section*{References}
Aerts, D. (1982). Description of many physical entities without the paradoxes encountered in quantum mechanics. {\it Foundations of Physics}, {\bf 12}, pp. 1131-1170.

\smallskip
\noindent
Aerts, D. (1983). Classical-theories and non-classical theories as a special case of a more general theory. {\it Journal of Mathematical Physics}, {\bf 24}, pp. 2441-2453.

\smallskip
\noindent
Aerts, D. (1996a). Framework for possible unification of quantum and relativity theories. {\it International Journal of Theoretical Physics}, {\bf 35}, pp. 2399-2416.

\smallskip
\noindent
Aerts, D. (1996b). Relativity theory: what is reality?. {\it Foundations of Physics}, {\bf 26}, pp. 1627-1644. 

\smallskip
\noindent
Aerts, D. (1999). The stuff the world is made of: physics and reality. In D. Aerts, J. Broekaert and E. Mathijs (Eds.), {\it Einstein meets Magritte: An Interdisciplinary Reflection} (pp. 129-183). Dordrecht: Springer. 

\smallskip
\noindent
Aerts, D. (2002). Being and change: foundations of a realistic operational formalism. In D. Aerts, M. Czachor and T. Durt (Eds.), {\it Probing the Structure of Quantum Mechanics: Nonlinearity, Nonlocality, Probability and Axiomatics} (pp. 71-110). Singapore: World Scientific. 

\smallskip
\noindent
Aerts, D. (2009a). Quantum structure in cognition. {\it Journal of Mathematical Psychology}, {\bf 53}, pp. 314-348.

\smallskip
\noindent
Aerts, D. (2009b). Quantum particles as conceptual entities: A possible explanatory framework for quantum theory. {\it Foundations of Science}, {\bf 14}, pp. 361-411.

\smallskip
\noindent
Aerts, D. (2010a). Interpreting quantum particles as conceptual entities. {\it International Journal of Theoretical Physics}, {\bf 49}, pp. 2950-2970.

\smallskip
\noindent
Aerts, D. (2010b). A potentiality and conceptuality interpretation of quantum physics. {\it Philosophica}, {\bf 83}, pp. 15-52.

\smallskip
\noindent
Aerts, D. (2011). Measuring meaning on the World-Wide Web. In D. Aerts, J. Broekaert, B. D'Hooghe and N. Note (Eds.), {\it Worldviews, Science and Us: Bridging Knowledge and Its Implications for Our Perspectives of the World}. Singapore: World Scientific. 

\smallskip
\noindent
Aerts, D. and Aerts, S. (1995). Applications of quantum statistics in psychological studies of decision processes. {\it Foundations of Science}, {\bf 1}, pp. 85-97.

\smallskip
\noindent
Aerts, D., Aerts, S. and Gabora, L. (2009). Experimental evidence for quantum structure in cognition. In P. D. Bruza, D. Sofge, W. Lawless, C. J. van Rijsbergen and M. Klusch (Eds.), {\it Proceedings of QI 2009-Third International Symposium on Quantum Interaction, Book series: Lecture Notes in Computer Science}, {\bf 5494}, pp. 59-70. Berlin, Heidelberg: Springer.

\smallskip
\noindent
Aerts, D., Broekaert, J. and Smets, S. (1999a). The liar paradox in a quantum mechanical perspective. {\it Foundations of Science}, {\bf 4}, pp. 115-132.

\smallskip
\noindent
Aerts, D., Broekaert, J. and Smets, S. (1999b). A quantum structure description of the liar paradox. {\it International Journal of Theoretical Physics}, {\bf 38}, pp. 3231-3239. 

\smallskip
\noindent
Aerts, D. and D'Hooghe, B. (2009). Classical logical versus quantum conceptual thought: Examples in economics, decision theory and concept theory. In P. D. Bruza, D. Sofge, W. Lawless, C. J. van Rijsbergen and M. Klusch (Eds.), {\it Proceedings of QI 2009-Third International Symposium on Quantum Interaction, Book series: Lecture Notes in Computer Science}, {\bf 5494}, pp. 128-142.

\smallskip
\noindent
Aerts, D. and Czachor, M. (2004). Quantum aspects of semantic analysis and symbolic artificial intelligence. {\it Journal of Physics A-Mathematical and General}, {\bf 37}, L123-L32.

\smallskip
\noindent
Aerts, D., Czachor, M., D'Hooghe, B. and Sozzo, S. (2010). The Pet-Fish problem on the World-Wide Web. {\it Proceedings of the AAAI Fall Symposium (FS-10-08), Quantum Informatics for Cognitive, Social, and Semantic Processes}, pp. 17-21.

\smallskip
\noindent
Aerts, D., Czachor, M., D'Hooghe, B. and Sozzo, S. (2010). The Pet-Fish problem on the World-Wide Web. {\it Proceedings of the AAAI Fall Symposium (FS-10-08), Quantum Informatics for Cognitive, Social, and Semantic Processes}, pp. 17-21. 

\smallskip
\noindent
Aerts, D. and Gabora, L. (2005a). A theory of concepts and their combinations I: The structure of the sets of contexts and properties. {\it Kybernetes}, {\bf 34}, pp. 167-191.

\smallskip
\noindent
Aerts, D. and Gabora, L. (2005b). A theory of concepts and their combinations II: A Hilbert space representation. {\it Kybernetes}, {\bf 34}, pp. 192-221.

\smallskip
\noindent
Aerts, D., Gabora, L., Sozzo, S. and Veloz, T. (2011). Quantum interaction approach in cognition, artificial intelligence and robotics. In {\it Proceedings of the Fifth International Conference on Quantum, Nano and Micro Technologies (ICQNM 2011)}, Nice, France, August 21-27, 2011.

\smallskip
\noindent
Arafat, S. and van Rijsbergen, C. J. (2007). Quantum theory and the nature of search. {\it Proceedings of the AAAI
Quantum Interaction Symposium}, pp. 114-122.

\smallskip
\noindent
Barrett, L. F., Tugade, M. M. and Engle, R. W. (2004). Individual differences in working memory capacity and dual-process theories of the mind. {\it Psychological Bulletin}, {\bf 130}, pp. 553-573.

\smallskip
\noindent
Blei, D. M., Ng, A. N. and Jordan, M. I. (2003). Latent Dirichlet Allocation. {\it Journal of Machine Learning Research}, {\bf 3}, pp. 993-1022.

\smallskip
\noindent
Bruner, J. (1990). {\it Acts of Meaning}. Cambridge, MA: Harvard University Press.

\smallskip
\noindent
Bruza, P. D. and Cole, R. J. (2005). Quantum logic of semantic space: An explanatory investigation of context effects in practical reasoning. In S. Artemov, H. Barringer, A. S. d'Avila Garcez, L. C. Lamb and J. Woods (Eds.), {\it We Will Show Them: Essays in Honour of Dob Gabbay}. London: College Publications.

\smallskip
\noindent
Bruza, P. D., Kitto, K., McEvoy, D. and McEvoy, C. (2008). Entangling words and meaning. In {\it Proceedings of the Second Quantum Interaction Symposium}, Oxford, UK, Oxford University Press, pp. 118-124.

\smallskip
\noindent
Bruza, P. D., Kitto, K., Nelson, D. and McEvoy, C. (2009). Extracting spooky-activation-at-a-distance from considerations of entanglement. In P. D. Bruza, D. Sofge, W. Lawless, C. J. van Rijsbergen and M. Klusch, (Eds.),{\it Proceedings of QI 2009-Third International Symposium on Quantum Interaction, Lecture Notes in Computer Science}, {\bf 5494}, Berlin, Heidelberg: Springer, pp. 71-83.

\smallskip
\noindent
Busemeyer, J. R., Wang, Z. and Townsend, J. T. (2006). Quantum dynamics of human decision-making. {\it Journal of Mathematical Psychology}, {\bf 50}, pp. 220-241.

\smallskip
\noindent
Busemeyer, J. R., Pothos, E., Franco, R. and Trueblood, J. S. (2011). A quantum theoretical explanation for probability judgment `errors'. {\it Psychological Review}, {\bf 108}, pp. 193-218.

\smallskip
\noindent
Deerwester, S., Dumais, S. T., Furnas, G. W., Landauer, T. K. and Harshman, R. (1990). Indexing by Latent Semantic Analysis. {\it Journal of the American Society for Information Science}, {\bf 41}, pp. 391-407.

\smallskip
\noindent
Freud, S. (1899). {\it Die Traumdeutung}. Berlin: Fischer-Taschenbuch.

\smallskip
\noindent
Gabora, L. and Aerts, D. (2002). Contextualizing concepts using a mathematical generalization of the quantum formalism. {\it Journal of Experimental and Theoretical Artificial Intelligence}, {\bf 14}, pp. 327-358.

\smallskip
\noindent 
Griths, T. L. and Steyvers, M. (2002). Prediction and semantic association. In {\it Advances in Neural Information Processing Systems}, {\bf 15}, pp. 11-18. Massachusetts: MIT Press.

\smallskip
\noindent
Hofmann, T. (1999). Probabilistic Latent Semantic Analysis. {\it Proceedings of the 22nd annual international ACM SIGIR conference on Research and development in information retrieval}, Berkeley, California, pp. 50-57.

\smallskip
\noindent
James, W. (1910). Some Problems of Philosophy. Cambridge, MA: Harvard University Press

\smallskip
\noindent
Kahneman, D. (2003). A perspective on judgment and choice: Mapping bounded rationality. {\it American Psychologist}, {\bf 58}, pp. 697-720. 

\smallskip
\noindent
Khrennikov, A. Y. and Haven, E. (2009). Quantum mechanics and violations of the Sure-Thing Principle: The use of probability interference and other concepts. {\it Journal of Mathematical Psychology}, {\bf 53}, pp. 378-388.

\smallskip
\noindent
Landauer, T. K. and Dumais, S. T. (1997). A solution of Plato's problem: The Latent Semantic Analysis theory
of the acquisition, induction, and representation of knowledge. {\it Psychological Review}, {\bf 104}, pp. 211-240.

\smallskip
\noindent
Landauer, T. K., Foltz, P. W. and Laham, D. (1998). Introduction to Latent Semantic Analysis. {\it Discourse
Processes}, {\bf 25}, pp. 259-284.

\smallskip
\noindent
Lund, K. and Burgess, C. (1996). Producing high-dimensional semantic spaces from lexical co-occurrence. {\it Behavior Research Methods, Instruments and Computers}, {\bf 28}, pp. 203-208.

\smallskip
\noindent
Paivio, A. (2007). {\it Mind and Its Evolution: A Dual Coding Theoretical Approach}. Mahwah, NJ: Lawrence Erlbaum Associates.

\smallskip
\noindent
Piron, C. (1976). {\it Foundations of Quantum Mechanics}. W. A. Benjamin: Reading.

\smallskip
\noindent
Piron, C. (1990). {\it M\'ecanique quantique bases et applications}. Presses Polytechniques et Universitaires Romandes: Lausanne.

\smallskip
\noindent
Pothos, E. M. and Busemeyer, J. R. (2009). A quantum probability explanation for violations of `rational' decision theory. {\it Proceedings of the Royal Society B}.

\smallskip
\noindent
Salton, G., Wong, A. and Yang, C. S. (1975). A vector space model for automatic indexing. {\it Communications of the ACM}, {\bf 18}, pp. 613-620.

\smallskip
\noindent
Sloman, S. A. (1996). The empirical case for two systems of reasoning. {\it Psychological Bulletin}, {\bf 119}, pp. 3-22.

\smallskip
\noindent
Sun, R. (2002). {\it Duality of the Mind}. Mahwah, NJ: Lawrence Erlbaum Associates.

\smallskip
\noindent
Van Rijsbergen, K. (2004). {\it The Geometry of Information Retrieval}, Cambridge, UK: Cambridge University Press.

\smallskip
\noindent
Widdows, D. (2003). Orthogonal negation in vector spaces for modeling word-meanings and document retrieval. In {\it Proceedings of the 41st Annual Meeting of the Association for Computational Linguistics}, pp. 136-143.

\smallskip
\noindent
Widdows, D. and Peters, S. (2003). Word vectors and quantum logic: Experiments with negation and disjunction. In {\it Mathematics of Language}, {\bf 8}, Indiana, IN: Bloomington, pp. 141-154.

\smallskip
\noindent
Widdows, D. (2003). Orthogonal negation in vector spaces for modelling word-meanings and document retrieval.
In Proceedings of the 41st Annual Meeting of the Association for Computational Linguistics (136-143). Sapporo,
Japan, July 7-12.

\smallskip
\noindent
Widdows, D. (2006). {\it Geometry and Meaning}. CSLI Publications: University of Chicago Press.

\smallskip
\noindent
Widdows, D. (2008). Semantic vector products: Some initial investigations. {\it Proceedings of the Second AAAI
Symposium on Quantum Interaction}. London: College Publications.

\smallskip
\noindent
Widdows, D. (2009). Semantic vector combinations and the synoptic gospels. {\it Lecture Notes In Articial Intelligence}, {\bf 5494}, pp. 251-265.

\end{document}